\documentclass[prl,twocolumn,a4paper,showpacs,floatfix,superscriptaddress]{revtex4} % double column

\usepackage{graphicx} %\newcommand{\figspace}{\vspace*{2ex}}
\usepackage{amsmath}
\usepackage{bm}
\usepackage{amssymb}

\begin{document}
\draft

%\twocolumn[ \hsize\textwidth\columnwidth\hsize\csname
%@twocolumnfalse\endcsname

\title{Exciton storage in a nano-scale Aharonov-Bohm ring with electric field tuning}

\author{Andrea M. Fischer}\affiliation{Department of Physics and Centre for Scientific Computing, University of Warwick, Coventry, CV4 7AL, United Kingdom}

\author{Vivaldo L.\ Campo Jr.}\affiliation{Departmento de Fisica, Universidade Federal de S\~{a}o Carlos-UFSCar, 13565-905 S\~{a}o Carlos, SP, Brazil}

\author{Mikhail E.\ Portnoi}\affiliation{School of Physics, University of Exeter, Exeter EX4 4QL, United Kingdom}

\author{Rudolf A.\ R\"{o}mer}\affiliation{Department of Physics and Centre for Scientific Computing, University of Warwick, Coventry, CV4 7AL, United Kingdom}

\date{$Revision: 1.40 $, compiled \today}

\begin{abstract}
We study analytically the optical properties of a simple model for an electron-hole pair on a ring subjected to perpendicular magnetic flux and in-plane electric field. We show how to tune this excitonic system from optically active to optically dark as a function of these external fields. Our results offer a simple mechanism for exciton storage and read-out.
\end{abstract}


\maketitle

%%%%%%%%%%%%%%%%%%%%%%%%%%%%%%%%%%%%%%%%%%%%%%%%%%%%%%%%%%%%%%%%%%%%%%%%
%\section{Introduction}
\label{sec-introduction}
%%%%%%%%%%%%%%%%%%%%%%%%%%%%%%%%%%%%%%%%%%%%%%%%%%%%%%%%%%%%%%%%%%%%%%%%
The manipulation of light and excitons is an area which has sparked much recent interest \cite{WinHB07,LunSLP99,But04,MitKM00,HauHDB99}. The speed of light has been slowed down to an amazing $17 \mathrm{ms}^{-1}$ in an ultracold gas of sodium atoms \cite{HauHDB99} by the use of electromagnetically induced transparency techniques \cite{HarFI90}. Bose-Einstein condensation of an excitonic gas is a phenomenon considered theoretically a long time ago \cite{KelK64}. However, it is only recently that the ability to grow high quality layered semiconductor structures has allowed for a successful experimental investigation of excitonic superfluids. Spatial patterns in photoluminescence measurements of an exciton gas in GaAs/AlGaAs coupled quantum wells have been observed \cite{ButGC02} and are thought to be a signature of quantum degeneracy \cite{LevSB05}. Zero Hall voltages measured in bilayer quantum-well systems for particular layer separations and magnitudes of a magnetic field applied perpendicular to the layers, are also understood to indicate the presence of an excitonic condensate \cite{KelEPW04}. There has also been experimental evidence for the formation of a polariton Bose-Einstein condensate \cite{KasRKB06} and a room temperature polariton laser \cite{ChrBGL07}. Such discoveries are very important for quantum computing applications \cite{DAmDBP02}. In current computers, electrons are used for information processing and photons for communication. Conversion between the two media places limitations upon the machine's efficiency. The creation of an exciton-based integrated circuit \cite{HigNBH08} could provide an exciting solution to this problem. The possibility of using excitons for data storage has also been investigated using indirect excitons in self-assembled quantum dots \cite{LunSLP99} and coupled quantum wells \cite{WinHB07}. Here the electron and hole are 
%spatially 
separated using a gate voltage, significantly prolonging their lifetimes. 

In this work we find a mechanism for controlling excitonic lifetimes in a single Aharonov-Bohm (AB) nanoring, again by the use of a constant electric field. Suppression of ground state exciton emission has already been seen for polarised excitons in the ring geometry for particular magnetic field strengths \cite{GovUKW02}; the introduction of impurities \cite{DiaUG04,DiaUS05} and the use of slightly eccentric rings \cite{GroZ07} makes previously dark states become optically active. 

\begin{figure}[tb]
  \centerline{
  \includegraphics[width=0.95\columnwidth]{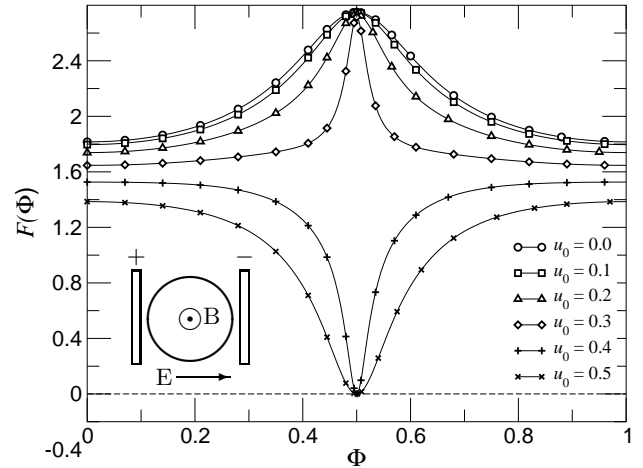}
%  \vspace{-10ex}
%  \hspace{-0.5\columnwidth}
%  \includegraphics[width=0.2\columnwidth]{fig-ringgeom.eps}
  }
  \begin{picture}(110,-32)(0,-70)
\thicklines
\put(5,-3){\circle{30}}
\thinlines
\put(5,-3){\circle{8}}
\put(5,-3){\circle*{2}}
\put(0,-25){\vector(1,0){20}}
\put(10,-5){$\mathrm{\bm{B}}$}
\put(-8,-28){$\mathrm{\bm{E}}$}
\put(-17,-20){\framebox(3,35){\mbox{ }}}
\put(24,-20){\framebox(3,35){\mbox{ }}}
\put(-19,18){$+$}
\put(22,18){$-$}
%\put(-20,-25){\framebox(50,50){\mbox{ }}}
  \end{picture}
  \caption{\label{fig-OS-v2} Oscillator strength $F$ as a function
    of $\Phi$ for electron-hole interaction strength $v_0=-2/\pi^2$ and various values of the electric field $u_0$. The horizontal dashed line indicates $\Delta=0$. Only every 6th data point is shown for clarity in each curve. Inset: Schematics of the ring geometry and the external fields.}
\end{figure}

In the present system, recombination of the electron-hole pair is not prevented by their confinement to different nanostructures, as previously seen, but by  appropriate tuning of external magnetic and electric fields. 
%
%Here we show how these effects can be produced by a simple sequence of changes in external electric and magnetic fields within a single device.
%The Aharonov-Bohm effect (ABE) for neutral excitons in semiconductor nanorings is very small and has not been experimentally detected yet. (?)
%
%%%%%%%%%%%%%%%%%%%%%%%%%%%%%%%%%%%%%%%%%%%%%%%%%%%%%%%%%%%%%%%%%%%%%%%%
%\section{Optical responses}
\label{sec-optics}
%%%%%%%%%%%%%%%%%%%%%%%%%%%%%%%%%%%%%%%%%%%%%%%%%%%%%%%%%%%%%%%%%%%%%%%%
In Fig.\ \ref{fig-OS-v2}, we show the oscillator strength $F$ of an exciton in a one-dimensional ring geometry threaded by a perpendicular magnetic flux $\Phi$ and in the presence of an in-plane electric field.
%-----------------------------------------------------------------------
%-----------------------------------------------------------------------
The oscillator strength oscillates as a function of magnetic flux. The most striking feature, however, is the way these oscillations change as the (dimensionless) electric field $u_0$ is varied. Up to a certain critical value of $u_0$, $F$ peaks at $\Phi=0.5$ (in units of the universal flux quantum $h/e$) indicating optimal conditions for the exciton creation and recombination. However, beyond this critical value, $F$ reaches zero at $\Phi=0.5$ and such transitions are forbidden. The critical electric field strength $u_{\rm c}$ is comparable to the electron-hole interaction strength $v_0$ and the transition occurs between $u_0=0.3$ and $u_0=0.4$ for the value used in Fig.\ \ref{fig-OS-v2}. This suggests a mechanism for controlling whether an excitonic state is optically active or not and thus whether an exciton is present or absent from the ring. The procedure is (i) begin with $\Phi=0$ and $u_0=0$ (ii) increase the flux adiabatically to $\Phi=0.5$ so an exciton can easily be created from the vacuum (iii) decrease $\Phi$ (iv) increase $u_0$ until it appreciably exceeds the critical value (v) increase flux to $\Phi=0.5$. The oscillator strength is now zero, so the exciton is trapped and unable to decay until the external fields are further tuned. This has important implications for excitonic data storage as we shall discuss later.

%An exact solution is found for the excitonic ground state energies, wavefunctions and corresponding oscillator strengths for the case of a contact electron-hole interaction. At low enough temperatures, the occupation of the ground state dominates the physics. 
%%%%%%%%%%%%%%%%%%%%%%%%%%%%%%%%%%%%%%%%%%%%%%%%%%%%%%%%%%%%%%%%%%%%%%%%
%\section{Single particle}
\label{sec-single-solution}
%%%%%%%%%%%%%%%%%%%%%%%%%%%%%%%%%%%%%%%%%%%%%%%%%%%%%%%%%%%%%%%%%%%%%%%%

We study the excitonic ground state energies, wavefunctions and corresponding oscillator strengths exactly for the case of a contact electron-hole interaction.
We obtain the excitonic wavefunction as a linear combination of the electron and hole single particle wavefunctions, $\psi^{\left( \mathrm{e}\right)}_N\left( \theta_\mathrm{e}\right)$ and $\psi^{\left( \mathrm{h}\right)}_{N'}\left( \theta_\mathrm{h}\right)$,
\begin{equation}
 \label{wf-exciton}
\Psi\left( \theta_\mathrm{e},\theta_\mathrm{h}\right) =\sum_{N,N'} A_{NN'}\psi^{\left( \mathrm{e}\right)}_N\left( \theta_\mathrm{e}\right)\psi^{\left( \mathrm{h}\right)}_{N'}\left( \theta_\mathrm{h}\right),
\end{equation}
depending on the azimuthal coordinates, $\theta_\mathrm{e}$ and $\theta_\mathrm{h}$, of the electron and hole respectively.
Here, the single particle electron wavefunction obeys the dimensionless Schr\"{o}dinger equation
\begin{eqnarray}
 \label{schro-elec}
 \lefteqn{\mathbf{H}(u_0,\Phi) \psi^{\left( \mathrm{e}\right)}\left( \theta_\mathrm{e}\right) 
 = 
 \lambda^{\left(\mathrm{e}\right)} \psi^{\left( \mathrm{e}\right)}
 \left( \theta_\mathrm{e}\right)} & & \\
& & \equiv \left[-\frac{d^2}{d{\theta_\mathrm{e}}^2}+2i\Phi\frac{d}{d\theta_\mathrm{e}}+\Phi^2-u_0\mathrm{cos}\left( \theta_\mathrm{e}\right)\right] \psi^{\left(\mathrm{e}\right)}\left( \theta_\mathrm{e}\right), \nonumber
\end{eqnarray}
where $\Phi=-\rho Ae/\hbar$, $u_0=e\rho U_0/\epsilon_0$, $\lambda^{\left( \mathrm{e}\right)} =E_\mathrm{e}/\epsilon_0$ and $\epsilon_0=\hbar^2/2m_\mathrm{e}\rho^2$ and $\rho$ is the ring radius, $A$ the magnetic vector potential, $U_0$ the constant in-plane electric field, $E_\mathrm{e}$ the energy of the electron and $m_\mathrm{e}$ the mass of the electron. The Schr\"{o}dinger equation is the same for the hole aside from a change in sign for $\Phi$ and $u_0$. The eigensolutions of \eqref{schro-elec} --- a Mathieu equation with external flux \cite{AS72} --- can be constructed by a Fourier expansion of the single-particle wavefunctions and obtaining the expansion coefficients and energies from the resulting matrix eigenequation.
%%%%%%%%%%%%%%%%%%%%%%%%%%%%%%%%%%%%%%%%%%%%%%%%%%%%%%%%%%%%%%%%%%%%%%%%
%\section
%\paragraph*{Self-consistent solution for the exciton}
\label{sec-exciton-solution}
%%%%%%%%%%%%%%%%%%%%%%%%%%%%%%%%%%%%%%%%%%%%%%%%%%%%%%%%%%%%%%%%%%%%%%%%

The Hamiltonian for the two particle excitonic system is given by $\bm{\mathrm{H}}(u_0,\Phi)+\mu \bm{\mathrm{H}}(-u_0/\mu,-\Phi)+V\left[R\left(\theta_\mathrm{e}-\theta_\mathrm{h} \right) \right]$,
%\begin{equation}
%\bm{\mathrm{H}}_\mathrm{e}=-\frac{d^2}{d{\theta_\mathrm{e}}^2}+2i\Phi\frac{d}{d\theta_\mathrm{%e}}+\Phi^2-u_0\mathrm{cos}\left( \theta_\mathrm{e}\right)
%\end{equation}
%is the single particle electron Hamiltonian,
%\begin{equation}
%\bm{\mathrm{H}}_\mathrm{h}=\mu \left( %-\frac{d^2}{d{\theta_\mathrm{h}}^2}-2i\Phi\frac{d}{d\theta_\mathrm{h}}+\Phi^2\right) %+u_0\mathrm{cos}\left( \theta_\mathrm{h}\right)
%\end{equation}
%is the single particle hole Hamiltonian, 
where $\mu=m_\mathrm{e}/m_\mathrm{h}$ is the ratio of the electron and hole masses \cite{mu} and $V$ is the interaction term, which depends on the distance $R$ between the electron and hole. We assume a short range interaction $V\left[ R\left( \theta_\mathrm{e}-\theta_\mathrm{h}\right)\right] /\epsilon_0=2\pi v_0\delta\left( \theta_\mathrm{e}-\theta_\mathrm{h}\right) $, where $v_0$ is the average interaction strength in units of $\epsilon_0$. $v_0$ is chosen as multiples of $-1/\pi^2$, so that the spatial extent of the exciton is comparable to the ring circumference, ensuring an appreciable AB effect \cite{Cha95}. The Schr\"odinger equation may now be written as
\begin{widetext}
\begin{equation}
 \label{schro-exciton}
\sum_{N,N'}A_{NN'}\left( \lambda_N^{\left( \mathrm{e}\right)}+\lambda_{N'}^{\left( \mathrm{h}\right)}-\Delta\right)\psi^{\left( \mathrm{e}\right)}_N\left( \theta_\mathrm{e}\right)\psi^{\left( \mathrm{h}\right)}_{N'}\left( \theta_\mathrm{h}\right)
 + 2\pi v_0\delta\left(\theta_\mathrm{e}-\theta_\mathrm{h} \right)\Psi\left( \theta_{\mathrm{e}},\theta_{\mathrm{h}}\right) =0   
\end{equation}
%\end{widetext}
where $\Delta$ is the excitonic energy in units of $\epsilon_0$. Multiplying by $\left[  \psi^{\left( \mathrm{e}\right)}_N\left( \theta_\mathrm{e}\right)\psi^{\left( \mathrm{h}\right)}_{N'}\left( \theta_\mathrm{h}\right)\right]  ^\dagger$ for particular $N,N' \in \mathbb{Z}$ and integrating over $\theta_{\mathrm{e}},\theta_{\mathrm{h}} \in \left[0,2\pi \right]$ allows us to obtain an expression for the coefficients in Eq.\ (\ref{wf-exciton})
%\begin{widetext}
\begin{equation}
 \label{exciton-coeffs}
A_{NN'}=-\frac{2\pi v_0}{\lambda_{N}^{\left( \mathrm{e}\right)}+\lambda_{N'}^{\left( \mathrm{h}\right)}-\Delta}
\int_0^{2\pi}d\theta \Psi\left( \theta,\theta \right){\psi^{\left( \mathrm{h}\right)}_{N'}\left( \theta\right)}^\dagger {\psi^{\left( \mathrm{e}\right)}_{N}\left( \theta\right)}^\dagger.
\end{equation}
%\end{widetext}
%Here we have used the property that the electron and hole wavefunctions are orthonormal. %This is due to the matrix in Eq.\ (\ref{eigval-elec}) being Hermitian and thus having orthogonal eigenvectors.(?) 
Multiplying Eq.\ (\ref{wf-exciton}) with $\theta_\mathrm{e}=\theta_\mathrm{h}=\theta$ by ${\psi^{\left( \mathrm{h}\right)}_{N'_0}\left( \theta\right)}^\dagger {\psi^{\left( \mathrm{e}\right)}_{N_0}\left( \theta\right)}^\dagger$ and integrating over $\theta \in \left[0,2\pi \right]$ gives
%\begin{widetext}
\begin{equation}
 \label{wf-exciton2}
\int_0^{2\pi}d\theta \Psi\left( \theta,\theta \right){\psi^{\left( \mathrm{h}\right)}_{N'_0}\left( \theta\right)}^\dagger {\psi^{\left( \mathrm{e}\right)}_{N_0}\left( \theta\right)}^\dagger =\sum_{N,N'} A_{NN'}\int_0^{2\pi}d\theta \psi^{\left( \mathrm{e}\right)}_N\left( \theta \right)\psi^{\left( \mathrm{h}\right)}_{N'}\left( \theta\right){\psi^{\left( \mathrm{h}\right)}_{N'_0}\left( \theta\right)}^\dagger {\psi^{\left( \mathrm{e}\right)}_{N_0}\left( \theta\right)}^\dagger.
\end{equation}
%\end{widetext}
Let 
\begin{equation}
\label{p}
P_{NN'N_0N'_0}=-\frac{2\pi v_0}{\lambda_{N}^{\left( \mathrm{e}\right)}+\lambda_{N'}^{\left( \mathrm{h}\right)}-\Delta} \int_0^{2\pi} d\theta \psi_N^{\left(\mathrm{e} \right) } \left( \theta \right)\psi_{N'}^{\left(\mathrm{h} \right)} \left( \theta \right){\psi_{N'_0}^{\left(\mathrm{h} \right) }}^\dagger \left( \theta \right){\psi_{N_0}^{\left(\mathrm{e} \right) }}^\dagger \left( \theta \right)
\end{equation}
and 
%\begin{widetext}
\end{widetext}
\begin{equation}
\label{g}
 G_{NN'}=\int_0^{2\pi}d\theta \Psi\left( \theta,\theta\right) {\psi_{N'}^{\left(\mathrm{h} \right) }}^\dagger \left( \theta \right){\psi_N^{\left(\mathrm{e} \right) }}^\dagger \left( \theta \right).
\end{equation}
Then Eq.\ (\ref{wf-exciton2}) can be expressed as
\begin{equation}
\label{exciton-eigeneq4d}
 G_{N_0N'_0}=\sum_{N,N'}G_{NN'}P_{NN'N_0N'_0}\left(\Delta \right).
\end{equation}
No analytic solution of \eqref{exciton-eigeneq4d} is known for finite $v_0$ and $u_0$ \cite{Cha95}. In order to find approximate solutions to \eqref{exciton-eigeneq4d}, we cut-off the sums at a maximum value $N_\mathrm{max}$ for $N,N',N_0,N'_0$. We have chosen $N_\mathrm{max}=15$ for calculation of the excitonic energies and $N_\mathrm{max}=11$ for the oscillator strengths; we have tested that our results do not change appreciably for the range of $\Phi$, $u_0$ and $v_0$ considered here. Mapping $N,N'\to K$ and $N_0,N'_0\to K_0$ according to $K=\left( N-1\right)N_\mathrm{max}+N'$, allows Eq.\ (\ref{exciton-eigeneq4d}) to be reformulated as a standard matrix equation
\begin{equation}
\label{exciton-eigeneq2d}
 G_{K_0}=\sum_{K}G_{K}P_{KK_0}\left(\Delta \right).
\end{equation}
The excitonic energies may now be calculated numerically by determining the values of $\Delta$ which result in the matrix $P_{KK_0}$ having an eigenvalue equal to $1$. For given $\Delta$, the eigenstates can be found using \eqref{g}, \eqref{exciton-coeffs} and \eqref{wf-exciton}.
%For fixed values of $u_0$ and $\Phi$, we test a range of $\Delta$ values with %separation 0.01 until we find a pair such that for one, the ground state %eigenvalue of $P_{KK_0}$ is less than 1 and for the other, it is greater than %1. A crude approximation for the excitonic energy is found by linearly %interpolating between these two points. The process is repeated closer to this %approximate value using a smaller separation of 0.002 to obtain a set of %coordinates $\left\lbrace \Delta, P_{KK_0}\left(\Delta \right)  %\mathrm{eigenvalue}\right\rbrace$. We find a function that interpolates these %points using the Mathematica function Interpolation. A final solution for the %excitonic energy is determined using the Mathematica function FindRoot to see %where the interpolation function equals 1.

\begin{figure}[tb]
  \centering
  \includegraphics[width=0.95\columnwidth]{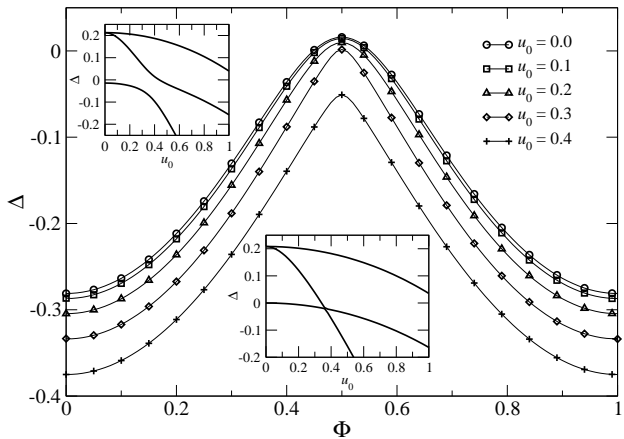}
  \caption{\label{fig-ex-phi} Excitonic energy as a function of
    magnetic flux $\Phi$ at various values of electric field $u_0$ for
    interaction strength $v_0=-2/\pi^2$. Only every 4th data point is shown for clarity in each curve. Inset top: The energies of the first $3$ states at $\Phi=0.45$ as a function of $u_0$. Inset bottom: The first $3$ states at $\Phi=0.5$ as a function of $u_0$. }
\end{figure}

\begin{figure}[tb]
  \centering
  \includegraphics[width=0.95\columnwidth]{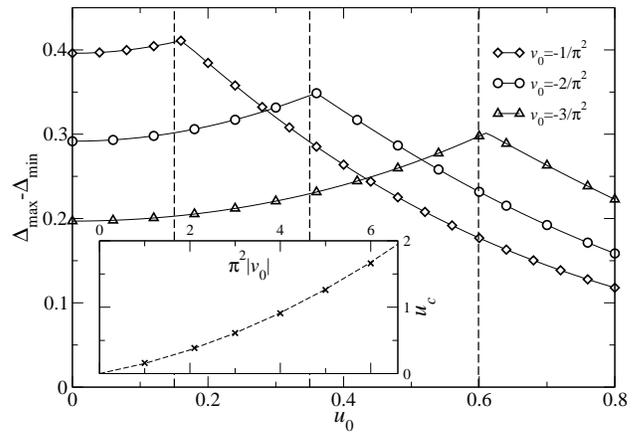}
  \caption{\label{fig-osc-amp} Amplitude of the Aharonov-Bohm 
    oscillations as a function of external field $u_0$ at various values of the interaction strength $v_0$. Only every 6th data point is shown for clarity in each curve. 
%Viv
Inset:  critical external field $u_{\rm c}$ as a function of $v_0$. The dashed curve, $u_{\rm c} = \pi |v_0| (\pi |v_0|/4 +2/5)$, fits the numerical points quite
well. 
%Viv
The vertical dashed lines in the main plot indicate $u_{\rm c}(v_0)$. 
}
\end{figure}

%%%%%%%%%%%%%%%%%%%%%%%%%%%%%%%%%%%%%%%%%%%%%%%%%%%%%%%%%%%%%%%%%%%%%%%%
%\section{Numerical results for the exciton}
\label{sec-results}
%%%%%%%%%%%%%%%%%%%%%%%%%%%%%%%%%%%%%%%%%%%%%%%%%%%%%%%%%%%%%%%%%%%%%%%%
Fig.\ \ref{fig-ex-phi} shows the resulting excitonic energies plotted as a function of magnetic flux for different electric field strengths. 
%-----------------------------------------------------------------------
%-----------------------------------------------------------------------
For small enough finite electric fields, the excitonic energy oscillates as a function of the magnetic field as seen previously for the zero electric field case \cite{Cha95}. This is due to the electron and hole, which were created simultaneously at the same position, having a finite probability to tunnel in opposite directions and meet each other on the opposite side of the ring. The dependence of the oscillation amplitude upon electric field strength is shown in Fig.\ \ref{fig-osc-amp} for different values of the interaction strength. 
%-----------------------------------------------------------------------
%-----------------------------------------------------------------------
In all cases there is initially a small increase \cite{MasC03} in amplitude $\Delta(\Phi=0.5)-\Delta(\Phi=0.0)$ and then it decreases strongly as a function of electric field strength. 
%Viv
This change happens when the polarization energy $2u_0$ becomes comparable to the exciton binding energy, which is $\pi^2 v_0^2/2$ on a line. As seen in the inset of Fig.~\ref{fig-osc-amp}, the polarization energy should be calculated with an effective electric field, $E_{\rm eff}$, where $eE_{\rm eff}R/\epsilon_0 = u_0 - 2\pi v_0/5$.
%Viv 
%This change happens at $u_0$ values comparable to the interaction strength, i.e.\ %when $|v_0| \approx \left\langle |u_0 \mathrm{cos} \theta|\right\rangle %=2|u_0|/\pi$. 
%
More precisely, the change of behaviour corresponds to a level crossing between ground and first excited state which only occurs at $\Phi=0.5$ as shown in the insets of Fig.\ \ref{fig-ex-phi}. The level crossing is associated with an exact symmetry at $\Phi=0.5$ upon exchanging $\theta_{\rm e,h}\rightarrow-\theta_{\rm e,h}$.
%As can be seen by the vertical dashed lines in Fig.\ \ref{fig-osc-amp}, this %estimate holds well for $v_0=-1/\pi^2,-2/\pi^2$ and approximately for %$v_0=-3/\pi^2$. Fig.\ \ref{fig-ex-u-v2} shows the excitonic energy as a function %of electric field strength. 
%-----------------------------------------------------------------------
\begin{figure}[tb]
  \centering
  \includegraphics[width=0.95\columnwidth]{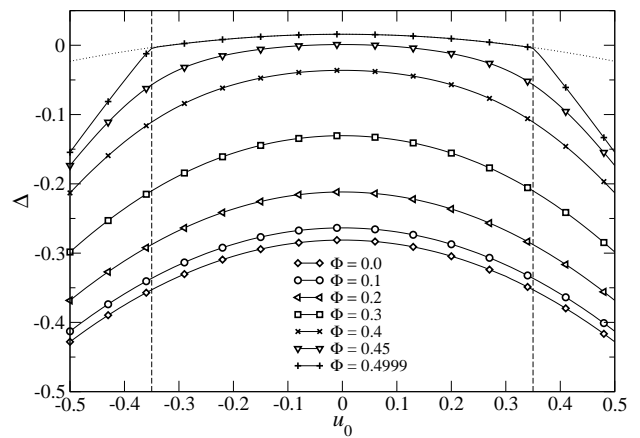}
  \caption{\label{fig-ex-u-v2} Excitonic energy as a function of $u_0$ for interaction strength $v_0=-2/\pi^2$ and various values of $\Phi$. Only every 6th data point is shown for clarity in each curve. The dotted lines show the results of second order perturbation theory performed on the ground state. The vertical dashed lines indicate $u_{\rm c}$.}
\end{figure}
%-----------------------------------------------------------------------
The quadratic behaviour for small enough magnetic fluxes and the negative curvature of $\Delta$ are as expected from second order perturbation theory in $u_0$ as shown in Fig.\ \ref{fig-ex-u-v2}. For magnetic flux values close to $\Phi=0.5$, the excitonic energy remains almost constant as a function of electric field until the critical $u_0$ value, when the level crossing occurs.
%electric field is comparable to the interaction strength, which is indicated by the vertical dashed lines. 
%Write more about Vivaldo here???
%Obviously, this behaviour indicates a higher-order exact cancellation of diagrams %in the $u_0$-perturbation series for $\Phi=0.5$.

\begin{figure}[tb]
 % \centering
  \centerline{
  (a)\includegraphics[width=0.45\columnwidth]{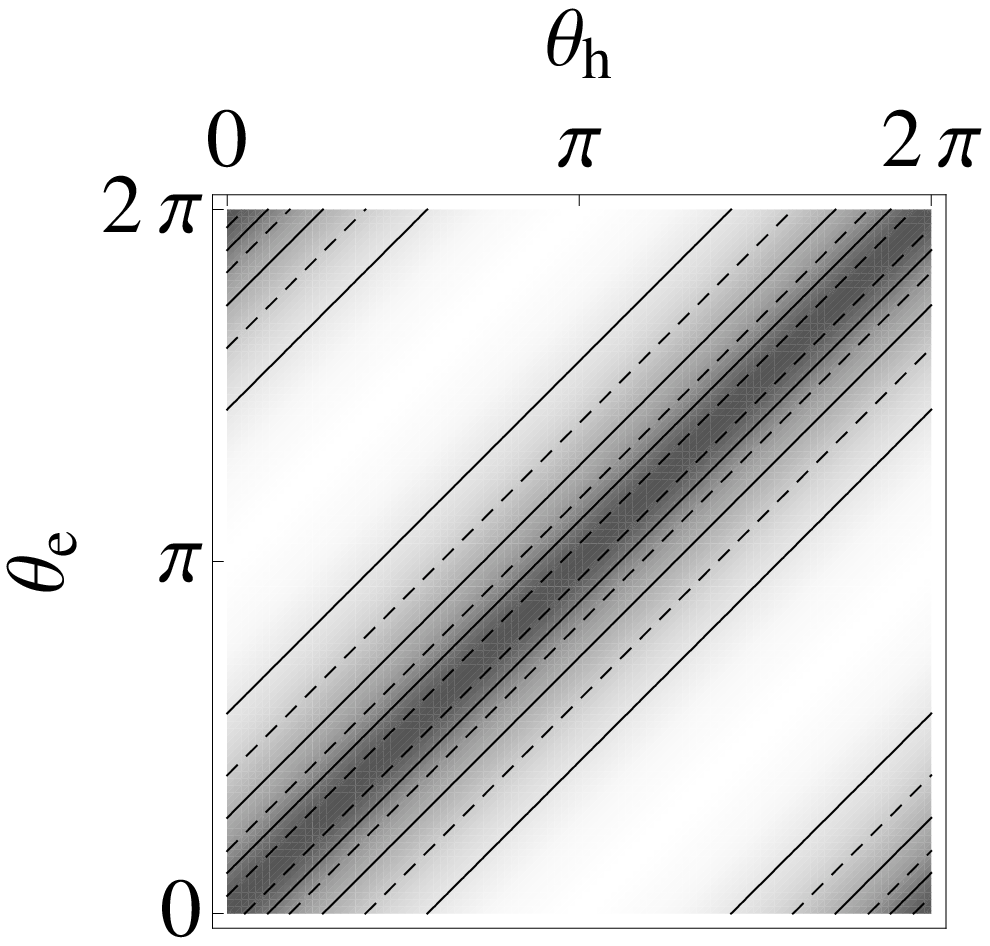}  %\includegraphics[width=0.05\columnwidth]{fig-legend.eps}}
  %\centerline{
  (b)\includegraphics[width=0.45\columnwidth]{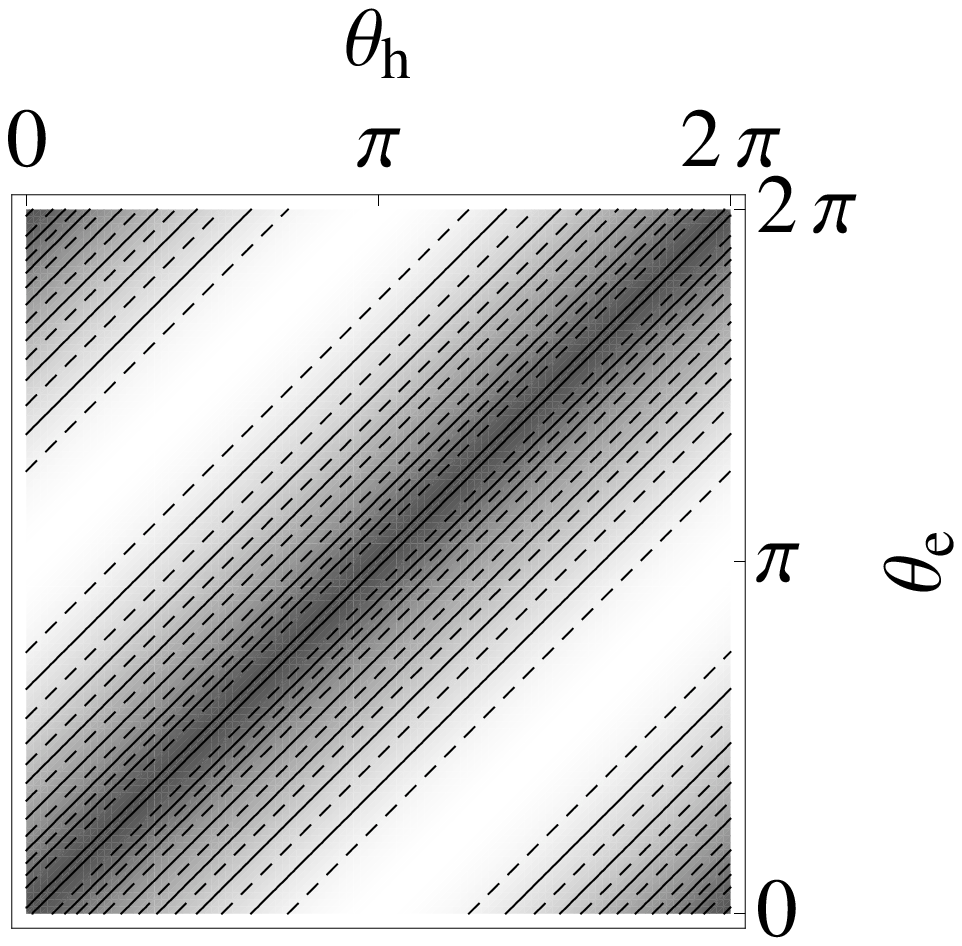}}
  \centerline{  
  (c)\includegraphics[width=0.45\columnwidth]{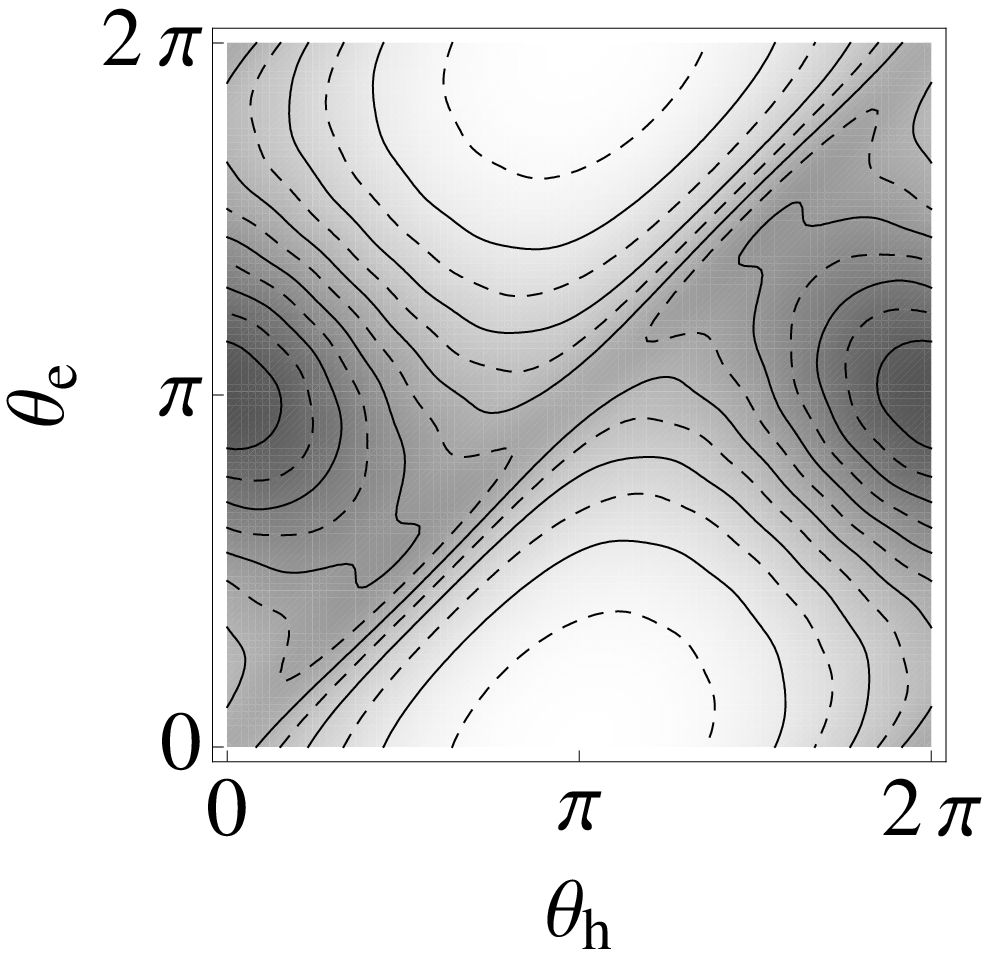}
  (d)\includegraphics[width=0.45\columnwidth]{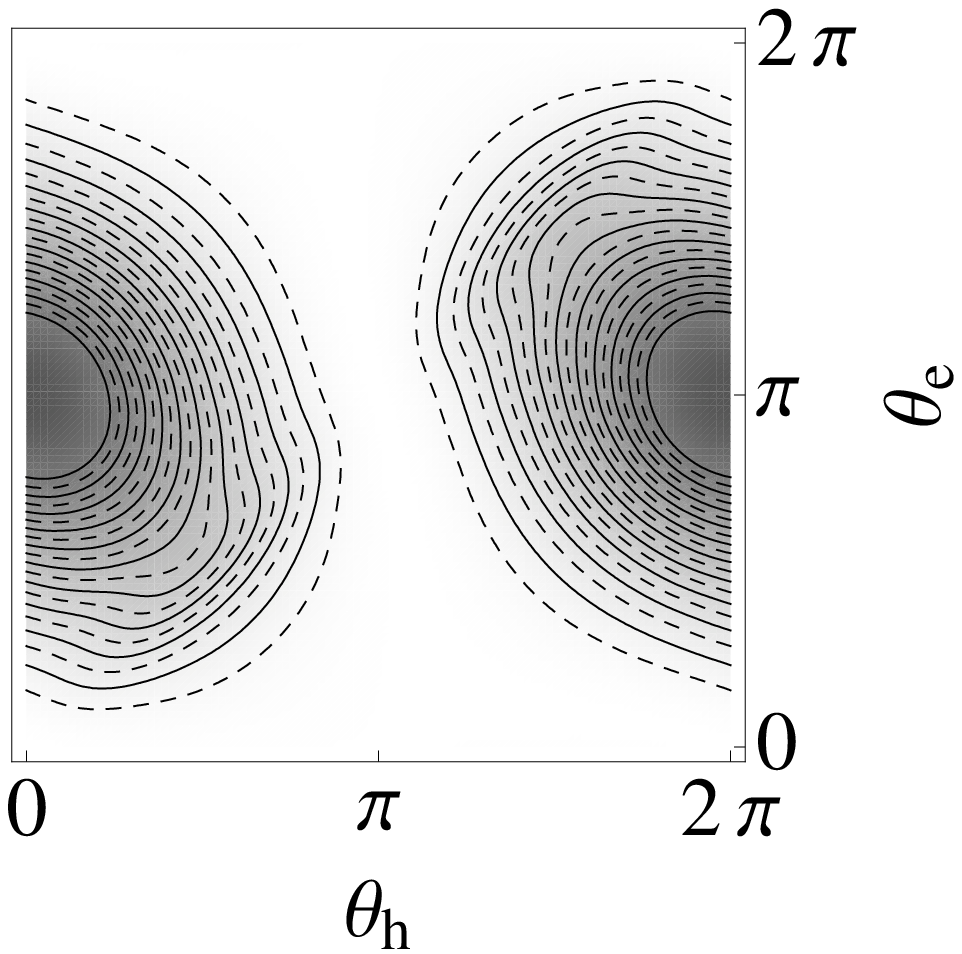}
  }
  \caption{
    Excitonic probability density
    $|\Psi(\theta_e,\theta_h)|^2$ at $v_0=-2/\pi^2$ and
    (a) $u_0=0$,
    $\Phi=0$,
    (b) $u_0=0$,
    $\Phi=0.4999$,
    (c) $u_0=0.5$,
    $\Phi=0$, and
    (d) $u_0=0.5$,
    $\Phi=0.4999$. The solid and dashed lines are contour lines separated in height by $1/10\pi^2$. }
\label{fig-ex-prob}
\end{figure}

The oscillator strength $F(\Phi)$ was already shown in Fig.\ \ref{fig-OS-v2} as a function of magnetic flux $\Phi$ for various values of electric field strength. It was determined according to
$F={|\int_0^{2\pi} d\theta\Psi\left( \theta,\theta\right)|^2}/{\int_0^{2\pi}d\theta_\mathrm{e}\int_0^{2\pi}d\theta_\mathrm{h}|\Psi\left( \theta_\mathrm{e},\theta_\mathrm{h}\right)|^2}
$,
%\begin{equation}
%\label{os}
%F=\dfrac{|\int_0^{2\pi} d\theta\Psi\left( %\theta,\theta\right)|^2}{\int_0^{2\pi}d\theta_\mathrm{e}\int_0^{2\pi}d\theta_\mathr%m{h}|\Psi\left( \theta_\mathrm{e},\theta_\mathrm{h}\right)|^2},
%\end{equation}
where $\Psi\left( \theta_\mathrm{e},\theta_\mathrm{h}\right)$ is the ground state excitonic wavefunction calculated as before. It measures the strength of a transition from the state $\Psi\left( \theta_\mathrm{e},\theta_\mathrm{h}\right)$ into the vacuum state. We note that the transition from optically active to dark coincides with the change in slope of the amplitude of the excitonic AB oscillation as in Fig.\ \ref{fig-osc-amp} and also the reinstatement of the parabolic behavior in Fig.\ \ref{fig-ex-u-v2}. In order to explain the mechanism for this behavior, we show in Fig.\ \ref{fig-ex-prob} the probability density function for the exciton at different values of $\Phi$, $u_0$ and $v_0$.
%-----------------------------------------------------------------------
In Fig.\ \ref{fig-ex-prob}(a) the magnetic flux and electric field have been set to zero. The dark grey diagonal region indicates a peak in probability density where the electron and hole coordinates are close together, so the exciton is very small. In Fig.\ \ref{fig-ex-prob}(b) the flux has been increased from its zero value, whilst keeping the interaction strength and electric field constant. There is a slight narrowing of the peaked region and the contours are closer together, showing that the exciton has become smaller. In Fig.\ \ref{fig-ex-prob}(c) the magnetic flux and interaction strength are the same as in Fig.\ \ref{fig-ex-prob}(a), but the electric field has been increased. We see the dark grey region has begun to split in two, suggesting the exciton is beginning to separate. In Fig.\ \ref{fig-ex-prob}(d) the magnetic flux has been increased from its zero value in Fig.\ \ref{fig-ex-prob}(c). The result is two dark grey peaks indicating that the exciton has been broken \cite{LunSLP99} into an electron and hole, which are located on opposite sides of the ring.
%In Fig.\ \ref{fig-ex-prob}(d) the electric field strength has been increased further and the two small dark grey regions corresponding to high peaks in the probability density, show that the exciton has been broken into an electron and hole, which are located on opposite sides of the ring. In Fig.\ \ref{fig-ex-prob}(b) the flux has been increased from its zero value in Fig.\ \ref{fig-ex-prob}(a), whilst keeping the interaction strength and electric field constant. There is a slight narrowing of the peaked region and the contours are closer together, showing that the exciton has become smaller. 
%In Fig.\ \ref{fig-ex-prob}(e) the magnetic flux has been increased from its zero value in Fig.\ \ref{fig-ex-prob}(d). The result is two dark grey peaks with steeper contour lines than in Fig.\ \ref{fig-ex-prob}(d) showing further separation of the electron and hole.
%%%%%%%%%%%%%%%%%%%%%%%%%%%%%%%%%%%%%%%%%%%%%%%%%%%%%%%%%%%%%%%%%%%%%%%%
%\section{Perturbative analysis at $\Phi=1/2$}
\label{sec-perturbation}
%%%%%%%%%%%%%%%%%%%%%%%%%%%%%%%%%%%%%%%%%%%%%%%%%%%%%%%%%%%%%%%%%%%%%%%%

%%%%%%%%%%%%%%%%%%%%%%%%%%%%%%%%%%%%%%%%%%%%%%%%%%%%%%%%%%%%%%%%%%%%%%%%
%\section{Conclusions}
\label{sec-conclusions}
%%%%%%%%%%%%%%%%%%%%%%%%%%%%%%%%%%%%%%%%%%%%%%%%%%%%%%%%%%%%%%%%%%%%%%%%

In conclusion, we have presented a simple model which allows the tuning of an exciton from light to dark by the application of external fields similar to previous experiments in quantum dots \cite{LunSLP99}. Rings of the required size have been fabricated before \cite{LorLFK99}.
The important feature is that the overall sensitivity can be optimized with the external magnetic field, while the control can be achieved with simple local electric gates. This suggests that our model might be useful in the quest to trap and store light \cite{WinHB07,LunSLP99,HauHDB99,HigNBH08}. Also, the contrast $F(u_0=0)/F(u_0 \gg 0)$ at $\Phi=0.5$ remains much more pronounced in larger rings than that of the ground state energy AB oscillations. Hence we expect that the effect is visible in somewhat larger rings and in the presence of weak impurities. 

\acknowledgements
We gratefully acknowledge discussions with M.\ Bergin, A.\ Chaplik, A.\ Pennycuick, G.\ Rowlands and S.\ A.\ Wells. AMF and RAR thank EPSRC for funding. MEP, RAR and VLC are grateful for hospitality at the International Centre for Condensed Matter Physics (Brasilia) where part of this work was done.

%%%%%%%%%%%%%%%%%%%%%%%%%%%%%%%%%%%%%%%%%%%%%%%%%%%%%%%%%%%%%%%%%%%%%%%%
% references
%%%%%%%%%%%%%%%%%%%%%%%%%%%%%%%%%%%%%%%%%%%%%%%%%%%%%%%%%%%%%%%%%%%%%%%%

%\bibliographystyle{prsty}\bibliography{bibliography/bibliograph}

\begin{thebibliography}{10}
\bibitem{mu} We restrict our presentation to $\mu=1$, but we have checked the validity of our results up to $\mu=1/10$. In particular the oscillator strength remains zero at $\Phi=0.5$ and large $u_0$.
\end{thebibliography}

%\begin{thebibliography}{99}
%\bibitem{mu} We restrict our presentation to $\mu=1$, but we have checked the %validity of our results up to $\mu=1/10$. In particular the oscillator strength %remains zero at $\Phi=0.5$ and large $u_0$.
%\bibitem{symmetry} Replacing $\Phi$ by $-\Phi$ in Eq.\ (\ref{eigval-hole}) produces a matrix similar to the one in Eq.\ (\ref{eigval-elec}). The similarity transformation is given by the matrix $\bm{\mathrm{S}}$ with entries $\mathrm{S_{ij}}=\left( -1\right)^{i+1}\delta_{ij}$.
%\end{thebibliography}

\end{document}